\documentclass[12pt,a4paper,final]{article}
 \usepackage[dvips,final]{graphicx}
  \usepackage[T2A]{fontenc}
    \usepackage[cp1251]{inputenc}
     \usepackage{bm}
      \usepackage{epsf}
       \usepackage{epsfig}
        \usepackage{amssymb,amsthm,amsmath}
         \usepackage{latexsym}
          \usepackage{pifont}
\graphicspath{{pictures/}}  
\usepackage{showkeys}   

\begin{document}
\begin{titlepage}
\begin{center}
~\\
~\\
~\\
~\\
\large \bfseries{THE DESCRIPTION OF SPATIAL CHARACTERISTICS OF ELASTIC PROCESSES IN THE WIGNER FUNCTION FORMALISM}
\mdseries
~\\
~\\
~\\
~\\
I. Perevalova\footnote[1]{IrenAdler1@rambler.ru},
М. Polyakov\footnote[2]{Maxim.Polyakov@tp2.rub.de},
O. Soldatenko\footnote[3]{o\_castor@mail.ru},
A. Vall\footnote[4]{anvall@mail.ru},

~

$^{1,3,4}$~\textit{- Physics Department, Irkutsk State University, Karl Marx str. 1, 664003, Irkutsk, Russia}\\~\\
$^{2}$~\textit{- Institut fur Theoretische Physik II Ruhr-Universitaet Bochum, NB6 D-44780 Bochum, Germany}

\end{center}
~
\begin{abstract}
The Wigner function formalism has been applied to the analysis of elastic scattering  processes. The new element of known formalism is the choice of the phase space on which the Wigner function is defined. This phase space is 4-dimensional space of canonically conjugate variables - the transverse momentum of scattered particle and the projection of the vector of the closest approach of particles in the plane which is perpendicular to the momentum of initial particle. The exact expression for the mean value of the scattered particles production area radius through a scattering amplitude has been obtained.
\end{abstract}

\end{titlepage}

~

\section{Introduction}
~~~~~Wigner function, introduced in 1932 \cite{Wigner}, defines quasiprobability distribution in phase space of canonically conjugate (coordinate-momentum) variables. The main object to construct the Wigner function is the wave function of the system. With the help of the Wigner function one can compute an average value of any dynamical observables.

The development of the formalism of the Wigner function considerably widened the field of its application. This formalism found an application in such fields as hydrodynamics \cite{Irving}, problem of N bodies in Boltzmann-Vlasov equation \cite{Carruthers}, plasma physics \cite{Brittin}. It was applied also to calculations of quantum corrections to the transition coefficients \cite{Choi}, as well as, to quantum optics \cite{Hillery, Scully, Schleich} and statistical mechanics \cite{Ford}. In Ref. \cite{Belitsky} the formalism of the Wigner function was applied to the quark and gluon distributions in the proton. In this work the relation of Wigner function to generalized proton distribution was demonstrated.

In the present work we generalize the formalism of the Wigner functions to the case of an elastic scattering of particles. We suggest to use, as a building block for a construction of the Wigner function, a scattering amplitude instead of a wave function of the system. In the case of the scattering amplitudes the connection between the momentum and coordinate space is not given by the Fourier transformation, as in the case of the standard definition of the Wigner function. The point is that the momrntum space in the case of scattering processes is not flat, the transformation from the momentum space to the canonically conjugate one is performed by the Shapiro function \cite{Shapiro}. We shall demonstrate that the canonically conjugate space space in the case of the elastic scattering corresponds to the space of the nearest approach parameter. Analogous formalism has been developed 
in Ref.~\cite{Kadyshevskii} in the contex of construction of the relativistic configuration space, as well as for the case of Wigner function in the Lobachevsky space \cite{Alonzo}.

The problem of construction of the canonically cojugate space for the case of the elasic scattering was solved in Ref.~\cite{PEPAN}. It was demosnstrated that the transformation from the momentum to the "coordinate" space is performed by the Shapiro functions which form the complete and orthonormal basis in the two dimensional space of the transverse momentum. The Shapiro functions are plane waves on the $SO(2.1)$ group.

\section{The $SO_{\mu}(2.1)$ algebra formalism}
~~~~~First we reproduce briefly the formalism of interaction area spatial structure description in the framework of the $SO_{\mu}(2.1)$ algebra. The formalism has been described in details in the papers \cite{PEPAN, Vall I, Vall II, EuroJ}. We consider here only elastic two-particle collisions.

A classical trajectory of free particle escaping from the interaction region continuated to the interaction region is located at certain time at a minimum distance from the $O$ point (the center of the target, Fig. 1). The coordinates of this point are described by the components of the vector $\vec{d}$. This vector is expressed through the particle momentum $\vec{q}$ and its orbital momentum $\vec{L}$ in the following way:
\begin{equation}\label{1}
d_i=\frac{1}{q^2}\varepsilon_{ijk}q_j L_k ~, ~~~~i,j,k=1,2,3~.
\end{equation}

\begin{figure}[h]
\begin{center}
\label{f:1}
\includegraphics*[scale=0.5] {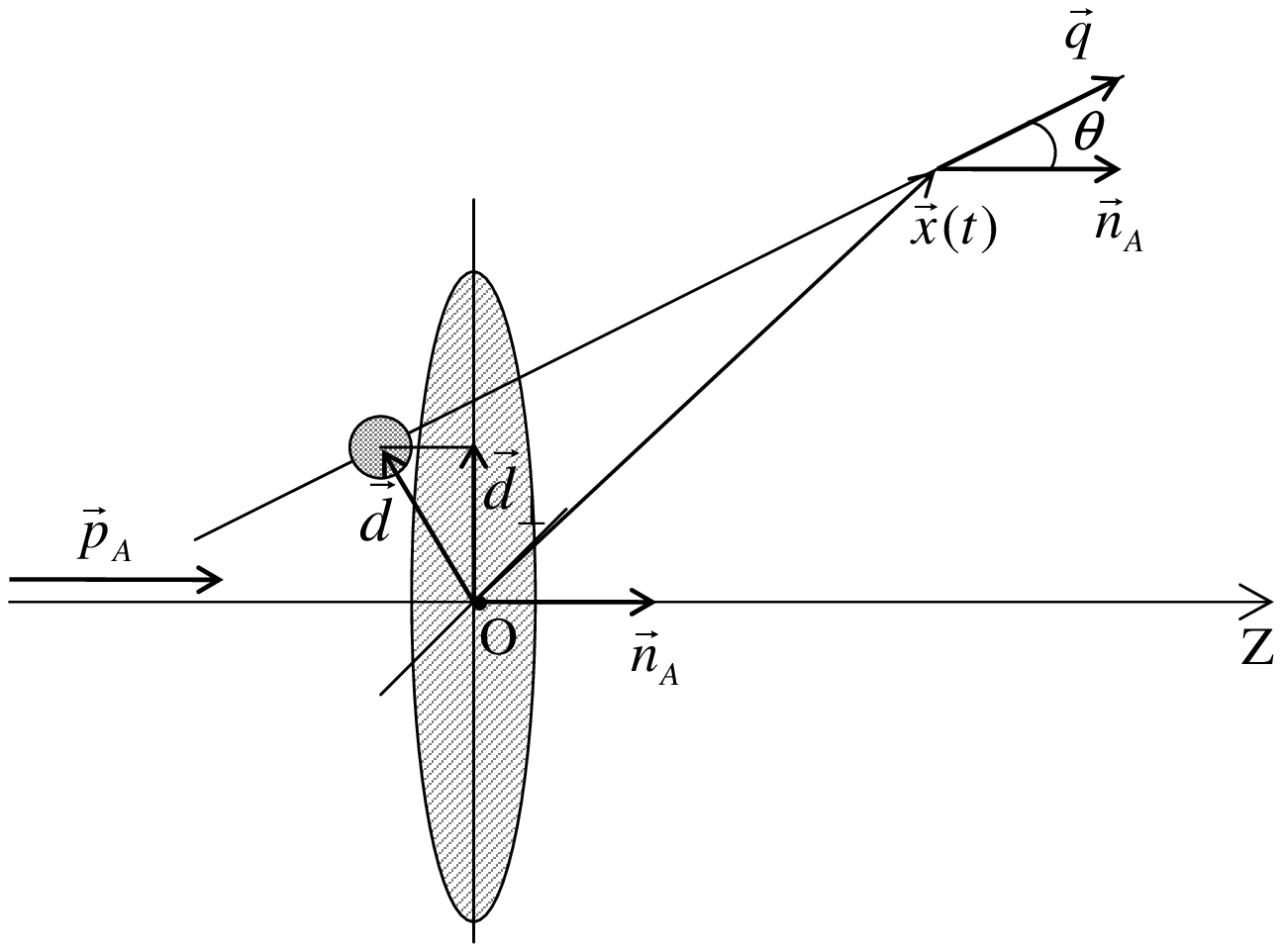}
\caption{\small Classical trajectory $\vec{X}(t)$ of the asymptotically free particle with the momentum $\vec{q}$ continued to the reaction
region is characterized by the minimum distance $\vec{d}$ from the chosen point $O$. This point corresponds to center of the target in the laboratory reference frame and to the beam collision point in the center-of-mass system.}
\end{center}
\end{figure}

When we consider two-particle elastic processes the vector $\vec{d}$ represents the vector of the closest approach. In the center-of-mass system it defines the mean minimum distance between scattered particles. In this case the $q_{i}$ in the ratio (\ref{1}) is relative momentum in the center-of-mass system and the $L_{i}$ is relative orbital momentum.

The next step is the quantization of $d_{i}$, it is based on the substitution of $C$-numbers by the corresponding operators in the expression (\ref{1}). The ordering of the operators in the expression (\ref{1}) is fixed by the requirement that the operator $d_i$ is Hermitian:
\begin{equation}\label{algebra1}
\begin{split}
d_{i}=\frac{1}{q^{2}}(\varepsilon_{ijk}q_{j}L_{k}-iq_{i})~,~~d_{i}=(d_{i})^{+}.
\end{split}
\end{equation}
(here and hereinafter $\hbar=1$).

Six generators $d_{1},~d_{2},~d_{3},~L_{1},~L_{2},~L_{3}$ form a $SO(3.1)$ algebra with the trivial Casimir operator. Operators $d_{1},~d_{2},~L_{3}$ constitute a nontrivial algebra $SO(2.1)$ with two Casimir operators $q^{2}$ and $\hat{K}$:
\begin{equation}\label{algebra2}
\begin{split}
[d_1 d_2]=-\frac{i}{q^2} L_3 ~,~~[d_1 L_3]=-i d_2 ~,~~[d_2 L_3]=i d_1~.
 \end{split}
\end{equation}
\begin{equation} \label{algebra3}
\begin{split}
\hat{K}=d^{~2}_{\perp}&-\frac{1}{q^2}L_3^2 \ \ ,\ \ d^{~2}_{\perp}=d_1^{~2}+d_2^{~2}\ \ ,\ [L_{3}\hat{K}]\neq 0~,\\ &[d_{1,2}~\hat{K}]=0~,~[d_{1,2}~q^2]=0~.
\end{split}
\end{equation}

Wave function of a state with fixed value of the Casimir operator and squared momentum satisfies the equations
\begin{equation}\label{2}
\left( d_{\perp}^{~2}-\frac{1}{q^2}~L_3^2 \right) \psi =b^{2} \psi ~,~~ \hat{q}^{~2} \psi = q^{2} \psi~.
\end{equation}

We interpret the eigenvalue $b^{2}$, which belongs continuous spectrum, as an escape parameter for a particle scattering process on the external field and as a squared vector of the closest approach in the case of two-particle elastic scattering.

Eq.~(\ref{2}) in the momentum representation on the surface $q^{2}=const$ are realized with the help of the following operators:
\begin{eqnarray}\label{oper_dekart-d3}
d_1&=& \frac{i}{q^2}\Big((q^2-q_1^2)\frac{\partial}{\partial q_1}-q_1 q_2\frac{\partial}{\partial q_2}-q_1\Big)~, \\
\nonumber
d_2&=& \frac{i}{q^2}\Big((q^2-q_2^2)\frac{\partial}{\partial q_2}-q_1 q_2\frac{\partial}{\partial
q_1}-q_2\Big)~,\\
\nonumber
L_3&=& -i\Big(q_1\frac{\partial}{\partial q_2}-q_2\frac{\partial}{\partial q_1}\Big)~.
\end{eqnarray}
Solutions of the system (\ref{2}) are the plane waves on the $SO(2.1)$ group \cite{Shapiro, Kadyshevskii, Fok}:
$$\psi(\vec{q}_{\bot})\equiv \xi(\vec{q}_{\bot},\vec{\mu}), $$
where the plane wave $\xi(\vec{q}_{\bot},\vec{\mu})$ is given by:
\begin{equation}\label{algebra34+}
\begin{split}
 \xi (\vec{q}_\bot ,{\vec{\mu}}) &=  \frac{q}{\sqrt{q^2-q^2_\bot
}}\left(\frac{q-\vec{n} \cdot \vec{q_\bot
}}{\sqrt{q^2-q^2_\bot}}\right)^{-\frac{1}{2}+i\mu}~,\\
&\vec{\mu}=\vec{n}\mu ~,~~ \mu^{2}=b^{2}q^{2}-\frac{1}{4}~,
\end{split}
\end{equation}
here the $\vec{n}$ is arbitrary unit vector. The set of functions $\xi(\vec{q}_{\bot},\vec{\mu})$ forms a complete ortonormal system. These functions allow us to perform transformation from the space of the functions defined in the momentum space $q_{\bot}$ to the space of the functions defined in the $\mu$-space, where $$\vec{q}_{\bot}=(q_{\bot}cos \varphi,~ q_{\bot}sin \varphi), ~0\leq q_{\bot}\leq q, ~0\leq \varphi \leq 2\pi,$$ $$\vec{\mu}=(\mu~cos \psi,~ \mu~sin \psi), ~0\leq \mu \leq \infty, ~0\leq \psi \leq 2\pi~.$$

Direct and inverse transformations between the function $f(\vec{q}_{\bot})$ in the momentum space and its image in the $\mu$-space $g(\vec{\mu})$ are given by:
\begin{equation}\label{matrel}
\begin{split}
&f^{(\epsilon)}(\vec{q}_{\bot})=\int\mathrm{d}\Omega_{\vec{\mu}}~\xi
(\vec{q}_\bot ,{\vec{\mu}})~g^{(\epsilon)}(\vec{\mu})~,\\
&g^{(\epsilon)}(\vec{\mu})=\frac{1}{(2\pi)^2}\int\mathrm{d}\Omega_{\vec{q}}~\bar{\xi}
(\vec{q}_\bot ,{\vec{\mu}})~f^{(\epsilon)}(\vec{q}_{\bot})~,
\end{split}
\end{equation}
where
\begin{equation*}\label{7+}
  \epsilon=\pm 1 ~,~~
  \mathrm{d}\Omega_{\vec{q}}=\frac{1}{q\sqrt{q^2-q^2_\bot }}\ \mathrm{d}\vec{q}_\bot\ ,\quad
  \mathrm{d}\Omega_{{\vec{\mu}}}=th(\pi\mu)\ \mathrm{d}{\vec{\mu}}~.
\end{equation*}

The signature $\epsilon$ correspondes to the sign of the third component of the momentum $q_{3}$. The transformations generated by the algebra (\ref{algebra2}) do not change the sign of third component of the momentum $q_{3}$. Therefore, division of the momentum space into $q_{3}> 0$ and $q_{3}< 0$ subspaces is invariant under transformations generated by  the $d_{1}~, ~d_{2}$ and $L_{3}$ operators. We introduce the $f^{(\pm)}$ amplitudes with fixed signature:
\begin{equation}
f^{(\pm)}(\vec{q}_{\bot})\equiv f(\vec{q}_{\bot},q_{3}=\pm \sqrt{q^{2}-q^{~2}_{\bot}})~. 
\end{equation}

We note that for arbitrary function $F(\vec{q})$ the following relation is satisfied:
\begin{equation}\label{Foc1}
  \int F(\vec{q})\  \mathrm{d}\vec{q}= \int q^2 dq \
  \mathrm{d}\Omega \ F(\vec{q})=\sum\limits _{\epsilon=\pm 1}\int q^2 dq \
  \mathrm{d}\Omega_{\vec{q}}\ F^{(\epsilon)}(\vec{q}_{\bot})~.
\end{equation}
Here
$$\mathrm{d}\Omega =sin \theta ~d \theta~ d\varphi \ \ , \ \mathrm{d}\Omega_{\vec{q}} = \frac{1}{q\sqrt{q^2-q^2_\bot }}\ \mathrm{d}\vec{q_\bot}~. $$

Further we introduce the expression for an elactic cross-section of a two-particle process in terms of elements of the $S$-matrix. As it is showed in Ref.~\cite{PEPAN}, elastic cross-section has the form
\begin{equation}\label{sigmaf}
\sigma^{(\pm)}_{el}(p)=\int
  \left|f^{(\pm)}(\vec{k}_{\bot};\vec{p})\right|^2~\mathrm{d}\Omega_{\vec{k}}~.
\end{equation}
Here the $\vec{p}$ ~ is the momentum of the incoming particle $A$ in the center-of-mass system, $\vec{k}$ is the momentum of scattered particle $A$, and $|\vec{p}|=|\vec{k}|=p$.

The amplitude $f^{(\pm)}(\vec{k}_{\bot};\vec{p})$ is related to the $S$-matrix as follows:
$$
f^{(\pm)}(\vec{k}_{\bot};\vec{p})=2\pi p~\lambda(p)~\langle\vec{k}_{\bot},\pm
\sqrt{k^2-k_{\bot}^2};\vec{k}_1=-\vec{k}|~A~|~\vec{p};-\vec{p}\rangle~,
$$
$$
\langle f|F|in\rangle=\delta^4(P_{in}-P_{f})\langle f|A|in\rangle,~~ S=I+iF~.
$$

The factor $\lambda(p)$ is related to the flux of colliding particles $|\vec{u}|$:
$$
\lambda(p)=\frac{1}{|\vec{u}|}=\frac{E_{A}\cdot E_{B}}{p(E_{A}+E_{B})}~.
$$ 

Detailed derivation of the relation (\ref{sigmaf}) has been given in Refs.~\cite{PEPAN, Shirkov}.

\section{The formalism of the $SO_{\mu}(2.1)$ group in the small transverse momentum approximation}

~~~~~We define the profile function $g^{(\epsilon)}(\vec{\mu})$ of two-particle state on the $SO_{\mu}(2.1)$ group for elastic processes as amplitude transform from the momentum space to the canonically conjugate $\vec{\mu}$-space, which is given by Eq.~(\ref{matrel}). This function plays a main role in the construction of the Wigner function. We will show that the formalism is simplified in the limit of small transverse momentum. This limit is justified in such processes where scattering occurs on small angles ($\theta \approx 0$ и $\theta \approx \pi$) into the forward or backward hemispheres. Both these cases correspond to the condition $q_{\bot}/q \ll 1$. It is useful to intriduce the momentum variables on the unit hyperboloid:
\begin{equation}
\begin{split}
u=(u_{0},~\vec{u})&=\left( \frac{q}{\sqrt{q^{2}-\vec{q}^{~2}_{\bot}}}~,~\frac{\vec{q}_{\bot}}{\sqrt{q^{2}-\vec{q}^{~2}_{\bot}}}\right)~,~~ \\
u^{2}&=u_{0}^{2}-\vec{u}^{~2}=1~.
\end{split}
\end{equation}

We have the folowing relations for these variables:
\begin{equation}
\begin{split}
\mathrm{d}\Omega_{\vec{q}}=\frac{\mathrm{d} \vec{u}}{u_{0}^{3}}~;~~~ \frac{\mathrm{d}\vec{q}_{\bot}}{q^{2}}&=\frac{\mathrm{d} \vec{u}}{u_{0}^{4}}~;~~~ \frac{\vec{q}_{\bot}}{q}=\frac{\vec{u}}{u_{0}}~;~~~ u_{0}=\sqrt{1+\vec{u}^{~2}}~;\\
\delta^{(2)}(\vec{u}-\vec{v})&=q^{2}\left( 1-\frac{k_{\bot}^{~2}}{q^{2}}\right)^{2} \delta^{(2)}(\vec{q_{\bot}}-\vec{k}_{\bot})~,~\\
\text{where}~v=(v_{0},~\vec{v})&=\left( \frac{q}{\sqrt{q^{2}-\vec{k}^{~2}_{\bot}}},~\frac{\vec{k}_{\bot}}{\sqrt{q^{2}-\vec{k}^{~2}_{\bot}}} \right)~.
\end{split}
\end{equation}

The equation (\ref{2}) for the eigenvalues of the Casimir operator $\hat{K}$ in the terms of these variables has the following form:
\begin{equation}\label{Ku}
(u_1^2+1)\frac{\partial^2 \psi}{\partial u_1^2}+(u_2^2+1)\frac{\partial^2\psi}{\partial u_2^2}+
2u_1 u_2\frac{\partial^2\psi}{\partial u_1\partial u_2} -\frac{2}{u_0^2} \psi+b^{2}q^{2}\psi =0~.
\end{equation}
Its solution (plane wave on the $SO_{\mu}(2.1)$ group) is given by:
\begin{equation}\label{plvolna}
\psi(u)=u_{0}\left( n\cdot u \right)^{-\frac{1}{2}+i\mu},
\end{equation}
\begin{equation}\label{virmu}
\begin{split}
n=(1, ~\vec{n})~,~~n^{2}&=0~,~~ n\cdot u=n_{0}u_{0}-\vec{n}\vec{u}=u_{0}-\vec{n}\vec{u}~,~~\\ \mu&=\sqrt{b^{2}q^{2}-\frac{1}{4}}~.
\end{split}
\end{equation}
(In the terms of momenta $\vec{q}$ this solution is presented in Eq.~(\ref{algebra34+})).

Now we consider the solution of the equation (\ref{Ku}) in the small transverse momenta approximation, that simplifies expression for $\psi(u)$. In turn it leads to essential simplification of expression for the Wigner function and analytic expressions for the profile function $g^{(\epsilon)}(\vec{\mu})$.
 
We look for a solution of Eq.~(\ref{Ku}) in the following form:
\begin{equation}\label{psi}
\psi(u)=e^{i \vec{\mu}\vec{u}}\varphi(u)~.
\end{equation}
Substituting this expression to the equation (\ref{Ku}), we obtain for the $\varphi(u)$ following differential equation:
\begin{equation}\label{urphi}
\begin{split}
(u_1^2+1)\left[ -\mu_{1}^{2} ~\varphi + 2i\mu_{1}\frac{\partial \varphi}{\partial u_1}+\frac{\partial^2\varphi}{\partial u_1^2}\right]+
(u_2^2+1)\left[ -\mu_{2}^{2} ~\varphi + 2i\mu_{2}\frac{\partial \varphi}{\partial u_2}+\frac{\partial^2\varphi}{\partial u_2^2}\right]+\\
+2u_1 u_2\left[ -\mu_{1}\mu_{2} ~\varphi + i\mu_{1}\frac{\partial \varphi}{\partial u_2}+i\mu_{2}\frac{\partial \varphi}{\partial u_1}+\frac{\partial^2\varphi}{\partial u_1\partial u_2}\right] -\frac{2}{1+\vec{u}^{~2}} ~\varphi+b^{2}q^{2}\varphi =0~.
\end{split}
\end{equation}

Our aim is to find $\varphi(u)$ in the region of $|\vec{u}| \ll 1$. In accordance with this we look for a solution for $\varphi(u)$ in the form of series over powers of the vector $\vec{u}$:
\begin{equation}\label{reshphi}
\varphi(u)=1+B_{ij}u_{i}u_{j}+O(u^{4})~,
\end{equation}
where the $B_{ij}$ is symmetric matrix. Substituting this expression to the equation (\ref{urphi}), we obtain the system:
\begin{equation}\label{syst}
\begin{split}
&\mu^{2}-b^{2}q^{2}+2-2Sp~B=0~,\\
&\mu_{1}B_{11}+\mu_{2}B_{12}=0~,\\
&\mu_{1}B_{12}+\mu_{2}B_{22}=0~.
\end{split}
\end{equation}
These expressions are obtained by equating coefficients in the (\ref{urphi}) in front of given power of the vector $u$. 

In what follows we shall require that the matrix $B=0$, i.e. the function $\varphi(u)$ in Eq.~(\ref{psi}) is unity with the accuracy of $O(u^{4})$. With that accuracy
\begin{equation}\label{exp1}
\psi(u)=e^{i \vec{\mu}\vec{u}}~.
\end{equation}
That implies that the transformation from the momentum space $\vec{u}$ to the canonically conjugate $\vec{\mu}$-space has simple form of the exponential function. The escape parameter squared $b^{2}$ in this case is the eigenvalue of the Casimir operator $\hat{K}$ with the accuracy of $O(u^{2})$:  
\begin{equation}\label{exp2}
\hat{K}e^{i \vec{\mu}\vec{u}}=b^{2}~e^{i \vec{\mu}\vec{u}}+O \left( \frac{(\vec{u} \cdot \vec{\mu})^{2}}{q^{2}}\right)~.
\end{equation}
The following conditions are satisfied: 
\begin{equation}\label{exp3}
b^{2}q^{2}-2=\mu^{2}~,~~ Sp~B=0~.
\end{equation}
Note that at this point we redefine the parameter $\mu$ as compared to Eq.~(\ref{virmu}). We make such redefinition in order to interpret the escape parameter $\vec{b}$ as the distance between scattered particles. 

In this approximation we can interpret amplitude $F^{(\pm)}(\vec{u})$ representation in the form
\begin{equation}\label{4}
F^{(\pm)}(\vec{u})=\int e^{i\vec{u}\vec{\mu}} g^{(\pm)}(\vec{\mu}) ~\mathrm{d}\vec{\mu}
\end{equation}
as expansion of the amplitude in eigenstates of Casimir operators (\ref{algebra3}), (\ref{2}) with corresponding interpretation of the vector $\vec{\mu}$. Inverse transform
\begin{equation}\label{psipribl1}
g^{(\pm)}(\vec{\mu})=\frac{1}{(2\pi)^{2}}\int e^{-i\vec{u}\vec{\mu}} F^{(\pm)}(\vec{u}) ~\mathrm{d}\vec{u}~
\end{equation}
defines the function $g^{(\pm)}(\vec{\mu})$. We call it as the profile function on the $SO_{\mu}(2.1)$ group in the small transverse momenta approximation. The momentum space area $\vec{u}=u(cos\varphi,~ sin\varphi),$ $0\leq u < \infty,~0\leq \varphi \leq 2\pi$, and the parameter space area $\vec{\mu}=\mu(cos\psi,~ sin\psi),~0\leq \mu < \infty, ~0\leq \psi \leq 2\pi$, are canonically conjugate to each other.

Approximate Eq.~(\ref{4}), in comparison to the exact equation (\ref{matrel}), essentially simplifies calculations and retains main features of the quantum character of the parameter $\vec{b}$: signature $\epsilon$, correct description of small phase spaces $bq$, i.e. region $\mu \approx 0$ and its physical interpretation as the vector of the closest approach.

Normalization of the amplitude $F^{(\pm)}(\vec{u})$ is chosen in such way that an elastic cross-section is given by:
\begin{equation}\label{5}
\sigma^{(\pm)}_{el}=\int \left| F^{(\pm)}(\vec{u})\right|^{2} ~\mathrm{d}\vec{u}~.
\end{equation}
Then the amplitude $F^{(\pm)}(\vec{u})$ is connected to the introduced above amplitude $f^{(\pm)}(\vec{q}_{\bot};\vec{p})$ by the following expression:
\begin{equation}\label{6}
F^{(\pm)}(\vec{u})=\frac{1}{(1+\vec{u}^{~2})^{3/4}}~f^{(\pm)}\left(\vec{q}_{\bot}=p\frac{\vec{u}}{\sqrt{1+\vec{u}^{~2}}};~\vec{p}_{\bot}=0,~p_{3}=p\right)~.
\end{equation}
Argument of the function $f^{(\pm)}$ on the right hand side of the (\ref{6}) is practically determinated by the substitution $|cos\theta|=\frac{1}{u_0}$. 

Expressions (\ref{4}) allow to construct Wigner function in standard way. The amplitude $F^{(\pm)}$ and the profile function $g^{(\pm)}$ play the role of the wave functions of the state in the momentum and coordinate spaces accordingly.

\section{Wigner function on the $SO_{\mu}(2.1)$ group}

~~~~~Let us introduce the $SO_{\mu}(2.1)$-transform of the correlator of scattering amplitudes:
\begin{equation}\label{w1}
W^{(\pm)}(\vec{u},\vec{\mu})=\frac{1}{(2\pi)^{2}}\int e^{i\vec{\mu} \vec{v}} F^{(\pm)} \left(\vec{u}-\frac{\vec{v}}{2}\right) ~\bar{F}^{(\pm)}\left(\vec{u}+\frac{\vec{v}}{2}\right) ~\mathrm{d}\vec{v}~,
\end{equation}
where the $F^{(\pm)}$ is amplitude determined by Eq.~(\ref{6}). This correlator is the generalisation of the Wigner function for the case of scattering amplitudes. In Eq.~(\ref{w1}) in contrast to usual definition of the Wigner function we use scattering amplitudes instead of wave functions. Introduced in that way function $W^{(\pm)}(\vec{u},\vec{\mu})$ has following properties:
\begin{equation}\label{w2}
\begin{split}
&1.~ \bar{W}^{(\pm)}=W^{(\pm)}~-~\text{(reality)}~,\\
&2.~ \int W^{(\pm)}(\vec{u},\vec{\mu}) ~\mathrm{d}\vec{\mu}=\left| F^{(\pm)}(\vec{u}) \right|^{2}=\frac{\mathrm{d}\sigma_{el}^{(\pm)}}{\mathrm{d}\vec{u}}~,\\
&3.~ \int W^{(\pm)}(\vec{u},\vec{\mu}) ~\mathrm{d}\vec{u}=(2\pi)^{2}\left|g^{(\pm)}(\vec{\mu})\right|^{2}=\frac{\mathrm{d}\sigma_{el}^{(\pm)}}{\mathrm{d}\vec{\mu}}~,~\text{ $g^{(\pm)}$ is determined by Eq.~(\ref{4})~,}\\
&4.~ \int W^{(\pm)}(\vec{u},\vec{\mu})~\mathrm{d}\vec{\mu} ~\mathrm{d}\vec{u}=\sigma^{(\pm)}_{el}~.
\end{split}
\end{equation}
Thus, this function satisfies all properties of the Wigner function and it allows one to calculate an average value of any function $h(\vec{u},~\vec{\mu})$ ordered according to the Weyl prescription \cite{Weyl}:
\begin{equation}\label{w3}
\left<h(\vec{u},~\vec{\mu})\right>^{(\pm)}=\frac{1}{\sigma_{el}^{(\pm)}} \int h(\vec{u},~\vec{\mu})~W^{(\pm)}(\vec{u},\vec{\mu})~\mathrm{d}\vec{\mu} ~\mathrm{d}\vec{u}~.
\end{equation}

Let us calculate the $\left<b^{2}\right>$ - average value of squared minimum distance between scattered particles in elastic collisions. In accordance with the (\ref{exp3}) this distance is directly connected with the $\left<\mu^{2}\right>$:
\begin{equation}\label{w33}
\left<\mu^{2}\right>=q^{2}\left<b^{2}\right>-2~. 
\end{equation}

It is follows from Eq.~(\ref{w3}) that
\begin{equation}\label{w4}
\left<\mu^{2}\right>^{(\pm)}=\frac{1}{\sigma_{el}^{(\pm)}} \int \mu^{2}~W^{(\pm)}(\vec{u},\vec{\mu})~\mathrm{d}\vec{\mu} ~\mathrm{d}\vec{u}~.
\end{equation}

Substituting here representation (\ref{w1}) and taking into account that 
$$
\frac{1}{(2\pi)^{2}}\int \mu^{2} e^{i\vec{\mu}\vec{u}} ~\mathrm{d}\vec{\mu}=-\nabla^{2}_{\vec{u}} ~\delta^{(2)}(\vec{u})~,
$$
we obtain
\begin{equation}\label{w5}
\left<\mu^{2}\right>^{(\pm)}=\frac{1}{\sigma_{el}^{(\pm)}} \int \left| \nabla_{\vec{u}} F^{(\pm)}(\vec{u}) \right|^{2}~\mathrm{d}\vec{u}~. 
\end{equation}
In the case when the amplitude $F^{(\pm)}(\vec{u})$ does not depend on the direction of vector $\vec{u}$, one has:
\begin{equation}\label{dop1}
\frac{\partial}{\partial u_{i}}F^{(\pm)}(\vec{u})=\frac{u_{i}}{u_{0}}~\frac{\partial F^{(\pm)}(\vec{u})}{\partial u_{0}}~, ~~i=1,~2~.
\end{equation}
Then representation (\ref{w5}) is reduced to the following one-dimensional integral:
\begin{equation}\label{dop2}
\left<\mu^{2}\right>^{(\pm)}=\frac{2\pi}{\sigma_{el}^{(\pm)}} \int \limits_{1}^{\infty} ~ (u_{0} - \frac{1}{u_{0}}) \left| \frac{\partial F^{(\pm)}(\vec{u})}{\partial u_{0}} \right|^{2}~du_{0}~.
\end{equation}
Analogous computations give 
\begin{equation}\label{w6}
\left<\mu_{k}\right>^{(\pm)}=\frac{1}{\sigma_{el}^{(\pm)}} \int \bar{F}^{(\pm)}(\vec{u})~\hat{\mu}_{k} ~F^{(\pm)}(\vec{u})~\mathrm{d}\vec{u}~,~~~~~ \hat{\mu}_{k}=i~\frac{\partial}{\partial u_{k}}~,~~~k=1,2.
\end{equation}
We note that the expression for $\left<\mu^{2}\right>^{(\pm)}$ can be put in the form similar to Eq.~(\ref{w6}):
\begin{equation}\label{w7}
\left<\mu^{2}\right>^{(\pm)}=\frac{1}{\sigma_{el}^{(\pm)}} \int \bar{F}^{(\pm)}(\vec{u})~\hat{\mu}^{2} ~F^{(\pm)}(\vec{u})~\mathrm{d}\vec{u}~.
\end{equation}
This expression corresponds to standard quantum-mechanical representations. Thus, the Wigner function allows to obtain dynamic variables depending on the $\vec{\mu}$ and $\vec{u}$ in the operator form and in any representation.

We emphasize that, althought the definition of the Wigner function (\ref{w1}) is an approximate\footnote{The approximation is related to the  choice of the approximate eigenfunction of the Casimir operator (\ref{exp1})}, Eqs.~(\ref{w5},\ref{dop2}) are exact in the frameworks of Wigner function formalism. The reason is  that the scattering amplitude drops fastly if $q_{\bot}\rightarrow q$ (angle $\theta \approx \pi/2$). Therefore the physical interpretation of the impact parameter $\mu$ retains even in the approximation we use.

Calculating $\left<b^{2}\right>$ it is important to reveal physical character of the parameter, i.e. its connection with dynamic parameters of a system. Let us consider simplest models for amplitudes - the one-particle exchange of a particle with the mass $m$ in the $t$-channel and the Regge amplitude with the Pomeron exchange.

The amplitude of one-particle exchange in the $t$-channel is:
\begin{equation}\label{wm1}
f(\vec{q},~\vec{p})=\frac{g}{t-m^{2}}=\frac{-g}{(\vec{p}-\vec{q})^{2}+m^{2}}~,~~|\vec{p}|=|\vec{q}|=p~.
\end{equation}
Here the $g$ is the coupling constant, $\vec{p}$ is the momentum of initial particle, $\vec{q}$ is the momentum of scattered particle.
Introducing signature and conical variables $u_{0}$ и $\vec{u}$, we obtain:
\begin{equation}\label{wm2}
f^{(\epsilon)}(\vec{q}_{\bot};~\vec{p}_{\bot}=0~,~p_{3}=p)=-\frac{g/2p^{2}}{1-\frac{\epsilon}{u_{0}}+\frac{m^{2}}{2p^{2}}}~,~~\epsilon=\pm 1~.
\end{equation}
From here we obtain 
\begin{equation}\label{wm3}
F^{(\epsilon)}(\vec{u})=-\frac{g}{2p^{2}}~\frac{u_{0}^{-3/2}}{1-\frac{\epsilon}{u_{0}}+\frac{m^{2}}{2p^{2}}}~.
\end{equation}
Using this expression, we obtain for the elastic cross-section
\begin{equation}\label{wm6}
\sigma^{(\epsilon)}_{el}=\frac{g^{2}}{4p^{4}}~\frac{2\pi}{z_{0}(z_{0}-\epsilon)}~,
\end{equation}
where
\begin{equation}\label{wm7}
z_{0}=1+\frac{m^{2}}{2p^{2}}~.
\end{equation}
(If $m=m_{A}+m_{C}$ parameter $z_{0}$ corresponds to the large half axis of Leman ellipse).

Substitution of the amplitude (\ref{wm3}) to the right hand side of Eq.~(\ref{dop2}) and taking into account Eq.~(\ref{wm6}) leads to the following result: 
\begin{equation}\label{wm8}
\left<\mu^{2}\right>^{(+)}=\frac{1}{4}~ z_{0}(z_{0}-1) \left(-\frac{2}{3}+\frac{4}{3(z_{0}-1)^2}+ 2z_{0}+4z_{0}^2+(4z_{0}^3-2z_{0})\ln\left(\frac{z_{0}-1}{z_{0}}\right)\right)~.
\end{equation}
In the region of large energies of colliding particles ($m^{2}/p^{2} \ll 1$) the second term in the brackets of Eq.~(\ref{wm8}) gives the main contribution. In this approximation we have:
\begin{equation}\label{wm9}
\sqrt{\left<b^{~2}\right>^{(+)}}=\sqrt{\frac{2}{3}}\cdot \frac{\hbar}{mc}~.
\end{equation}
In the case of scattering into backward hemisphere similar computations lead to following result:
\begin{equation}\label{wm10}
\left<\mu^{2}\right>^{(-)}=\frac{1}{4}~ z_{0}(z_{0}+1) \left(-\frac{2}{3}-2z_{0}+4z_{0}^2+\frac{4}{3(z_{0}+1)^2} +(2z_{0}-4z_{0}^3)\ln\left(\frac{z_{0}-1}{z_{0}}\right)\right)~.
\end{equation}
In the large energies region we obtain:
\begin{equation}\label{wm11}
\left<b^{~2}\right>^{(-)}=0.39~\frac{\hbar^{2}}{p^{2}}~.
\end{equation}

Thus, from Eq.~(\ref{wm9}) one can see that in the case of one-particle exchange average "impact" radius of scattered particles into forward hemisphere is determined by the Compton length of exchange particle (virtual "cloud"). It is a well known fact and its confirmation allows us to use the expression (\ref{wm9}) for calculations with more complicated  amplitude models $F(\vec{u})$. Elastic scattering into backward hemisphere (reflex scattering) is determined by the impact radius which drops with the increase of energy (\ref{wm11}). The existence of non-trivial backward scattering is a consequence of quantum character of the closest approach vector, and it is an essential distinction from eikonal models. As it has been shown in Eq.~\cite{EuroJ}, correct using unitarity condition leads to an important contribution of scattering into backward hemisphere.

As a second example let us consider the model of the elastic $pp$-scattering corresponding to exchange of the Pomeranchuk vacuum pole in the $t$-channel. In this case the amplitude of the elastic scattering is:
\begin{equation}\label{wmr1}
\begin{split}
&f(cos\theta)=ig(s)e^{Bt}~,\\
&B=\alpha^{'}(0)\left(-i\frac{\pi}{2}+\ln\frac{s}{s_{0}}+\varkappa\right)~,\\
&t=-2p^{2}(1-cos\theta)~.
\end{split}
\end{equation}

The phenomenological parameter $\varkappa$ determines slope of the diffractive cone in differential cross-section of the $pp$-scattering, $s$ as is invariant mass, $s_{0}\approx 100~GeV^{2}$ determines the border of the high-energy regime area. The parameter $\alpha^{'}(0)$ is the slope of the Pomeranchuk linear trajectory: $\alpha(t)=1+\alpha^{'}(0)t$. According to ISR and SPS experimental data at energies $\sqrt{s} \approx 540~GeV$ the parameter $\varkappa$ is $\varkappa=12 \div 16$,~~ $\alpha^{'}(0)=0.12\pm 0.03~GeV^{-2}$~\cite{Levin}.

Transformation to cone variables gives for the amplitude $F^{(\epsilon)}(\vec{u})$:
\begin{equation}\label{wmr2}
F^{(\epsilon)}(\vec{u})=\lambda e^{2p^{2}B\frac{\epsilon}{u_{0}}} \cdot u_{0}^{-3/2}~,~~~\lambda=ig(s)e^{-2p^{2}B}~.
\end{equation}
Taking into account Eq.~(\ref{5}) and the relation $\mathrm{d}\vec{u}=u_{0}\mathrm{d}u_{0}\mathrm{d}\varphi$, we obtain:
\begin{equation}\label{wmr3}
\sigma^{(\epsilon)}_{el}(p)=2\pi |\lambda|^{2} \int \limits_{1}^{\infty} \frac{\mathrm{d}u_{0}}{u_{0}^{2}}~ e^{a/u_{0}}= 2\pi |\lambda|^{2} ~\frac{e^{a}-1}{a}~,
\end{equation}
where $a=4p^{2}\epsilon (Re~B)$.
Elastic cross-section of scattering into forward hemisphere is:
\begin{equation}\label{wmr3_1}
\sigma^{(+)}_{el}(p)=\frac{\pi~g^{2}(s)}{2~p^{2}~ \alpha^{'}(0)~\left(\ln \frac{s}{s_{0}}+\varkappa\right)}~.
\end{equation}
Taking into account Eq.~(\ref{dop2}), we obtain for $\left<\mu^{2}\right>^{(\epsilon)}$:
\begin{equation}\label{wmr4}
\left<\mu^{2}\right>^{(\epsilon)}=\frac{2\pi}{\sigma^{(\epsilon)}_{el}} |\lambda|^{2}\int \limits_{1}^{\infty} \left(u_{0}-\frac{1}{u_{0}}\right)~ \left| \frac{\partial}{\partial u_{0}}\left(u_{0}^{-3/2}e^{\frac{2p^{2}\epsilon B}{u_{0}}}\right) \right|^{2}~.
\end{equation}
In the high energy approximation we have $|B| \gg 1$ and $a \gg 1$. In this approximation we obtain for the forward hemisphere:
\begin{equation}\label{wmr5}
\left<\mu^{2}\right>^{(+)}=2p^{2} \alpha^{'}(0)~ \frac{\pi^{2}/4+\left(\varkappa+\ln \frac{s}{s_{0}}\right)^{2}}{\varkappa+\ln \frac{s}{s_{0}}}~.
\end{equation}
In accordance with (\ref{w33}) the average value of the squared minimum distance between scattered particles is given by: 
\begin{equation}\label{wmr6}
\left<b^{2}\right>^{(+)}=2 \alpha^{'}(0)~ \frac{\pi^{2}/4+\left(\varkappa+\ln \frac{s}{s_{0}}\right)^{2}}{\varkappa+\ln \frac{s}{s_{0}}}~.
\end{equation}
This distance is determined by the parameter of diffractive cone slope $\varkappa$, the Pomeranchuk trajectory slope $\alpha^{'}(0)$ and by the energy factor $\ln s/s_{0}$ corresponding to Regge-exchange dynamic.

At $s\approx s_{0}$ the main contribution to $\left<b^{2}\right>^{(+)}$ is determined by the parameter $\varkappa$. For this case we have
\begin{equation}\label{wmr7_1}
\sqrt{\left<b^{2}\right>^{(+)}}\approx (0.34 ~\div~ 0.39)~fm~.
\end{equation}
At the SPS-collider energy $\sqrt{s} \approx 540~GeV$:
\begin{equation}\label{wmr7_2}
\sqrt{\left<b^{2}\right>^{(+)}}\approx (0.44 ~\div~ 0.48)~fm~.
\end{equation}
At the LHC energy $\sqrt{s} \approx 14~TeV$:
\begin{equation}\label{wmr7_3}
\sqrt{\left<b^{2}\right>^{(+)}}\approx (0.50 ~\div~ 0.54)~fm~.
\end{equation}
For reliable estimations of the $\left<b^{2}\right>^{(+)}$ it is necessary to use realistic amplitude $F(\vec{u})$ of the elastic process. The amplitude should be in agreement with contemporary experimental data and it should satisfy unitarity condition. This condition controls contribution of the many-particle intermediate states to the elastic process and it gives considerable contribution to $\left<b^{2}\right>^{(+)}$ as energy function. 

Further we calculate the Wigner function $W^{(+)}(\vec{u},~\vec{\mu})$ for the model of one-particle exchange (\ref{wm1}). For its calculation we should note that the amplitude of scattering into forward hemisphere (\ref{wm3}) is simplified in the following way:
\begin{equation}\label{K1-1}
F^{(+)}(\vec{u})=-\frac{g}{2p^{2}}~\frac{u_{0}^{-3/2}}{1-\frac{1}{u_{0}}+\frac{m^{2}}{2p^{2}}}~\approx \frac{-g/p^{2}}{\vec{u}^{~2}+\frac{m^{2}}{p^{2}}}~.
\end{equation}
In this case we obtain for the Wigner function:
\begin{equation}\label{K1-2}
\begin{split}
W^{(+)}(\vec{u},~\vec{\mu})=\frac{g^{2}}{(2\pi)^{2}~ p^{4}} \int e^{i\vec{\mu}\vec{\Delta}} \frac{\mathrm{d}\vec{\Delta}}{\left[ \left(\vec{u}+\frac{\vec{\Delta}}{2}\right)^{2}+\frac{m^{2}}{p^{2}}\right] \left[ \left(\vec{u}-\frac{\vec{\Delta}}{2}\right)^{2}+\frac{m^{2}}{p^{2}}\right]}~.
\end{split}
\end{equation}
Reality of the $W^{(+)}(\vec{u},~\vec{\mu})$ is ensured by the symmetry of the denominator under transformation $\vec{\Delta} \rightarrow -\vec{\Delta}$. For the calculation of the integral (\ref{K1-2}) we use the following expression:
\begin{equation}\label{K1-3}
\begin{split}
\frac{1}{\left[ \left(\vec{u}+\frac{\vec{\Delta}}{2}\right)^{2}+\frac{m^{2}}{p^{2}}\right] \left[ \left(\vec{u}-\frac{\vec{\Delta}}{2}\right)^{2}+\frac{m^{2}}{p^{2}}\right]}&= \\
&=\int \limits_{0}^{1}  \frac{\mathrm{d}\alpha}{\left[ \alpha \left(\vec{u}-\frac{\vec{\Delta}}{2}\right)^{2}+(1-\alpha)\left(\vec{u}+\frac{\vec{\Delta}}{2}\right)^{2}+\frac{m^{2}}{p^{2}}\right]^{2}}~.
\end{split}
\end{equation} 
Then we have:
\begin{equation}\label{K1-4}
\begin{split}
&W^{(+)}(\vec{u},~\vec{\mu})=\\
&=\frac{g^{2}}{(2\pi)^{2}~p^{4}} ~ Re\int \mathrm{d}\vec{\Delta}~ e^{i\vec{\mu}\vec{\Delta}}\int \limits_{0}^{1}  \frac{\mathrm{d}\alpha}{\left[ \alpha \left(\vec{u}-\frac{\vec{\Delta}}{2}\right)^{2}+(1-\alpha)\left(\vec{u}+\frac{\vec{\Delta}}{2}\right)^{2}+\frac{m^{2}}{p^{2}}\right]^{2}}~.
\end{split}
\end{equation} 
The operation $Re$ ensures the reality of the function $W^{(+)}(\vec{u},~\vec{\mu})$.
Further we make the following substitution:
\begin{equation}\label{K1-5}
\vec{\delta}=\frac{\vec{\Delta}}{2}+(1-2\alpha)\vec{u}~,
\end{equation}
and obtain:
\begin{equation}\label{K1-6}
\begin{split}
&W^{(+)}(\vec{u},~\vec{\mu})=\\
&=\frac{4~g^{2}}{(2\pi)^{2}~p^{4}}~Re \int \limits_{0}^{1} \mathrm{d}\alpha ~\int \mathrm{d}\vec{\delta}~e^{2i\vec{\mu}\vec{\delta}} e^{-2i\vec{\mu}\vec{u}(1-2\alpha)} \frac{1}{\left[\vec{\delta}^{~2}+\frac{m^{2}}{p^{2}}+4\alpha~\vec{u}^{~2}(1-\alpha)\right]^{2}}~.
\end{split}
\end{equation}
Using the expression
\begin{equation}\label{K1-7}
\frac{1}{\left(\delta^{2}+a^{2}\right)^{2}}=\int \limits_{0}^{\infty} \mathrm{d}t~t~e^{-t(\delta^{2}+a^{2})}~,
\end{equation}
where we introduce $a^{2}=m^{2}/p^{2}+4\alpha u^{2}(1-\alpha)$, we obtain:
\begin{equation}\label{K1-8}
\begin{split}
W^{(+)}(\vec{u},~\vec{\mu})&=\frac{4~g^{2}}{(2\pi)^{2}~p^{4}}~Re \int \limits_{0}^{1} \mathrm{d}\alpha~ e^{-2i\vec{\mu}\vec{u}(1-2\alpha)}~\int \limits_{0}^{\infty} \mathrm{d}t~t~e^{-ta^{2}}~\int \mathrm{d}\vec{\delta}~e^{2i\vec{\mu}\vec{\delta}-t \delta^{2}}=\\
&=\frac{4~g^{2}\pi}{(2\pi)^{2}~p^{4}}~Re \int \limits_{0}^{1} \mathrm{d}\alpha~ e^{-2i\vec{\mu}\vec{u}(1-2\alpha)}~\int \limits_{0}^{\infty} \mathrm{d}t~e^{-ta^{2}-\mu^{2}/t}~.
\end{split}
\end{equation}
The integral over variable $t$ is expressed through the cylindrical function \cite{Gradshtein}:
\begin{equation}\label{K1-82}
\int \limits_{0}^{\infty} \mathrm{d}x \cdot x^{\nu-1} e^{-\beta x-\gamma/x}=2\left(\frac{\gamma}{\beta}\right)^{\nu/2} K_{\nu}\left(2\sqrt{\beta \gamma}\right)~.
\end{equation}
Then
\begin{equation}\label{K1-10}
\begin{split}
&W^{(+)}(\vec{u},~\vec{\mu})=\\
&=\frac{4~g^{2}}{2\pi~p^{4}}~ \int \limits_{0}^{1} \mathrm{d}\alpha~ \frac{|\vec{\mu}|~cos\left[2\vec{\mu}\vec{u}(1-2\alpha)\right]}{\sqrt{4\alpha(1-\alpha)u^{2}+\frac{m^{2}}{p^{2}}}}~ K_{1}\left(2|\vec{\mu}|\sqrt{4\alpha(1-\alpha)u^{2}+\frac{m^{2}}{p^{2}}}\right)~.
\end{split}
\end{equation}
This expression is explicit form of the Wigner function for the model of the one-particle exchange. This function of the $|\vec{\mu}|$ и $|\vec{u}|$ at different angles $\varphi$ between these vectors is presented in Figs. 2 and 3 (with arbitrary normalisation). 

\begin{figure}[ht]
\begin{center}
\label{PicWig1}
\includegraphics*[scale=0.7] {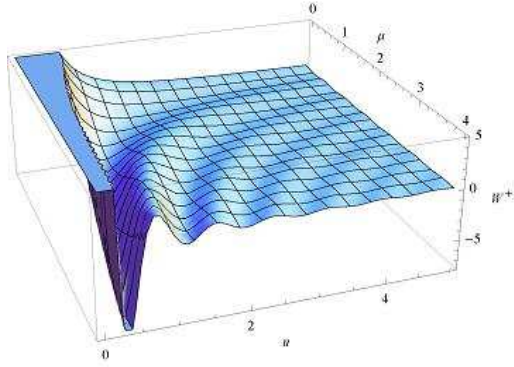}
\caption{\small The Wigner function for the model of the one-particle exchange. Dependence on $|\vec{\mu}|$ and $|\vec{u}|$ at the $\varphi=0$ is shown.}
\end{center}
\end{figure}

\begin{figure}[ht]
\begin{center}
\label{PicWig2}
\includegraphics*[scale=0.7] {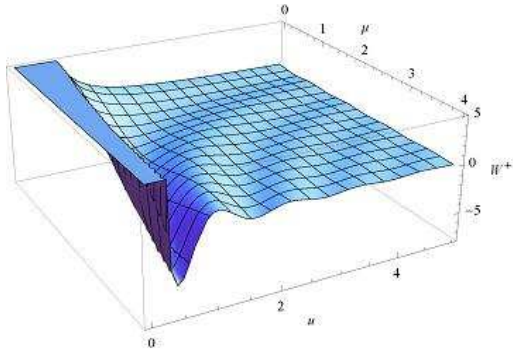}
\caption{\small The Wigner function for the model of the one-particle exchange. Dependence on $|\vec{\mu}|$ and $|\vec{u}|$ at the $\varphi=\pi/4$ is shown.}
\end{center}
\end{figure}
One can see from Eq.~(\ref{K1-10}) and numerical calculation that the Wigner function is sign-changing function. It is its usual property.

\section{Conclusion}

~~~In the present work we constructed the generalization of the Wigner function, using as the building blocks the elastic scattering amplitudes. The canonically conjugate (to the transverse momentum) space is formed by the vector of the nearest approach of the scattered particles. Constructed in this way the Wigner function allows one to compute space-time characteristics of the scattering processes such as the mean radius of the particle production region, multiplicities of produced particles with fixed production radius. The developed here approach also allows one to obtain the tomographic image of the particle production region with help of Radon transformation, this formalism we shall discuss elsewhere.

\section{Acknowledgements}
~~~The work was supported in parts by the Grant NSh-3810.2010.2 of the President of the Russian Federation for Support of Leading Scientific Schools, by the Analytical Departamental Special-Purpose Programme "Development of the Scientific Potencial of Higher School (2009-2010)", projects RNP.2.2.1.1/1483, RNP.2.2.1.1/1539, by the Russian Federal Programme
“Research and Teaching Experts in Innovative Russia”
the project 1.5 (contract 02.740.11.5154) and by the project 1.3.1 (NK-653P), and by the German Ministry for Education and Research
(grant 06BO9012).

\end{document}